\documentclass{aa}

\usepackage{graphicx}
\usepackage{caption}
\usepackage{subcaption}
\usepackage{amssymb}
\usepackage{amsfonts}
\usepackage[export]{adjustbox}
\usepackage[varg]{txfonts}
\usepackage[utf8]{inputenc}
\usepackage{amsmath}
\usepackage{xcolor}
\usepackage{hyperref}
\usepackage{comment}
\usepackage[normalem]{ulem}
\usepackage{upgreek}

\newcommand{\rstar}{\ensuremath{R_{*}}}

\newcommand{\Mns}{\ensuremath{M_{\rm *}}}
\newcommand{\epsBS}{\ensuremath{\epsilon_{\rm BS}}}
\newcommand{\betaBS}{\ensuremath{\beta_{\rm BS}}}

\newcommand{\ximag}{\ensuremath{\xi_{\rm mag}}}

\renewcommand{\Re}{\ensuremath{R_{\rm e}}}
\newcommand{\Rm}{\ensuremath{R_{\rm e}}}

\newcommand{\xib}{\ensuremath{\xi_{\rm B}}}

\newcommand{\del}[1]{}

\newcommand{\glaccept}[1]{#1}

\defcitealias{BS76}{BS}
\defcitealias{AL23}{AL}
\defcitealias{shock-free}{LAI}

\begin{document}

\title{Variable opacity and accretion-column regimes of magnetized neutron stars}

\titlerunning{Variable opacity and accretion-column regimes}

\author{
G. Lipunova\inst{1,2}
\and
P. Abolmasov\inst{3}
}

\institute{
    Dr.~Karl Remeis-Observatory and Erlangen Centre for Astroparticle Physics, Friedrich-Alexander Universit\"at Erlangen-N\"urnberg, Sternwartstr.~7, 96049 Bamberg, Germany \email{galina.lipunova@fau.de}
\and
Max-Planck-Institut f\"ur Radioastronomie, Auf dem H\"ugel 69, 53121 Bonn, Germany
\and
The Raymond and Beverly Sackler School of Physics and Astronomy, Tel Aviv University, Tel Aviv 69978, Israel
}

\abstract
{Radiation-supported accretion columns in magnetized neutron stars can operate in several distinct regimes. As the accretion rate increases, magnetospheric accretion may proceed through hot-spot emission, efficiently cooling radiative shocks, and advective sinking columns.}
{
The structure of the accretion flow is determined by the global parameters such as magnetic field strength and mass accretion rate, but is also affected by opacity variations in strong magnetic fields. 
A decrease of opacity for photons with energy below the cyclotron energy is able to substantially augment the classical critical accretion rate separating efficiently cooling solutions from advective solutions. }
{
We propose an approach to determine the critical accretion rate above which the Basko–Sunyaev sinking solution becomes unavoidable.
We also use the Basko--Sunyaev sinking solution to estimate the plasma temperature 
and compare the characteristic photon energy with the local cyclotron energy. 
}
{For a combination of moderate magnetic fields and high accretion rates, the characteristic photon energy in advective columns typically exceeds the cyclotron energy. 
In this case, the opacity is not  reduced by the magnetic field, and tall accretion columns are formed.
Strong magnetic-field systems may retain short, efficiently cooling accretion columns over a wide range of super-Eddington accretion rates. We identify a region in accretion rate and magnetic-field strength where both efficiently cooling and advective solutions may coexist. In this region, the column may switch or oscillate between the  regimes and geometries.}
{
The typical time of such oscillations are close to the replenishment time of
the column, or to the thermal time near its bottom.
The quasi-periodic oscillations, observed in some super-Eddington objects in
the 1-100~mHz frequency range, may be related to such relaxation cycles.} 
{}
\keywords{
Stars: neutron --
accretion, accretion discs --
X-rays: binaries --
radiative transfer --
magnetohydrodynamics (MHD)
}

\maketitle

\section{Introduction}

Accretion onto magnetized neutron stars (NS) is controlled by the interaction between the accretion flow and the stellar magnetic field. At sufficiently high accretion rates, radiation pressure becomes dynamically important and the flow forms a radiation-supported structure above the stellar surface. Depending on the accretion rate, geometry, and cooling efficiency, the flow may appear as a radiative shock close to the NS surface, an extended advective accretion column, or a fully subsonic shock-free magnetospheric flow~(see the accompanying paper by~\citealt{shock-free}, which we will hereafter refer to as \citetalias{shock-free}).

In the model of a radiation-supported accretion column developed by \citet{BS76} (hereafter \citetalias{BS76}), the accretion flow is stopped by radiation pressure above the NS surface. For moderate accretion rates, the released gravitational energy can escape through the column walls. At higher accretion rates, the photon diffusion time becomes too long, and part of the energy is advected downward with the flow. This gives rise to the sinking regime of \citetalias{BS76}. Time-dependent simulations by \citet{AL23}, hereafter \citetalias{AL23}, confirmed that advective columns can form and that mass and heat leakage through the column walls may regulate the structure of the flow.

A key uncertainty in this problem is the  opacity. In strong magnetic fields, the electron scattering cross-section depends on photon energy, polarization, propagation direction, and magnetic-field strength.
The opacity can be reduced below the Thomson value when the characteristic photon energy lies below the cyclotron energy. At high energies, cyclotron resonances and pair production tend to increase the opacity, whereas Klein–Nishina corrections reduce it. 
{In relativistic radiation magnetohydrodynamic simulations (RRMHD) of supercritical neutron star accretion columns,  \citet{Sheng+2023} found that using temperature-dependent magnetic
scattering polarization-averaged Rosseland opacities dramatically affected both the dynamics and the time-averaged structure of the accretion column.}

Generally, the opacity variations are important because the critical accretion rate separating efficiently cooling columns from advective columns is controlled by the photon escape time. A reduced opacity allows efficient cooling at higher accretion rates and therefore favours shorter accretion columns. If the opacity is close to the Thomson value, or enhanced, the column becomes advective more easily and can extend to a significant fraction of the magnetosphere.

In this paper we analyse how opacity variations below the cyclotron energy affect the classification of magnetospheric accretion regimes. In Sect.~\ref{sec.regimes}, we summarize the sequence of regimes and introduce the cooling/advection criterion. In Sect.~\ref{sec:opacity_bs}, we examine whether the Thomson-opacity approximation is self-consistent in advective sinking columns by comparing the characteristic photon energy with the local cyclotron energy.  
In Sect.~\ref{sec:opacity_regimes}, we estimate how the decrease in opacity  in a strong magnetic field shifts the boundaries between efficiently cooling and advective column solutions. We discuss the implications of column-height transitions for observations in Sect.~\ref{sec:discussion}, and summarize in Sect.~\ref{sec:conc}.

\section{Census of magnetospheric accretion regimes}\label{sec.regimes}

\begin{figure}
    \centering
\includegraphics[width=0.9\linewidth]{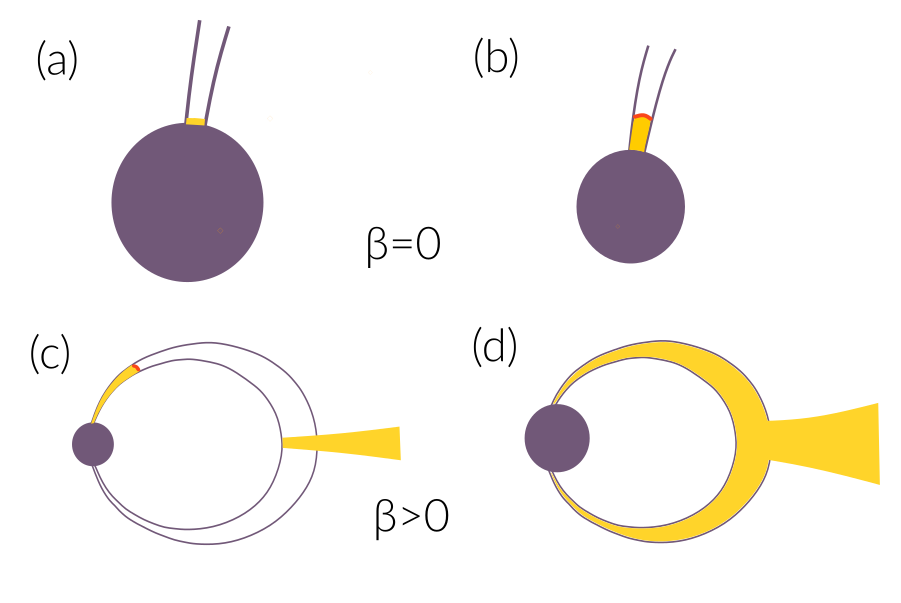}
   \caption{Magnetospheric regimes, ordered by increasing accretion rate: (a) hot spot, (b) efficiently cooling accretion column, (c) advective sinking regime (disc shown as well), and (d) shock-free magnetospheric accretion. In the two last cases, the solution to the conservation equations is characterized by a non-zero falling velocity at the NS surface, thereby indicating a flow of heat at the base of the accretion column (advection parameter $\beta>0$, see Sect.~\ref{sec.regimes}). 
   }
       \label{fig:drawing}
\end{figure}

Fig.~\ref{fig:drawing} schematically illustrates four regimes of magnetospheric accretion onto a magnetized NS. At low accretion rates, the infalling plasma is stopped in the NS atmosphere, producing a hot spot on the stellar surface (panel a). At higher accretion rates, radiation pressure becomes dynamically important and a radiative shock forms at some distance above the surface. Below the shock, the plasma is decelerated to a low velocity and sinks slowly towards the NS.

The post-shock flow may then operate in different regimes depending on the efficiency of radiative cooling. If the released gravitational energy is efficiently radiated through the column walls, a relatively short, efficiently cooling accretion column is formed (panel b). If cooling is inefficient, part of the energy is advected with the sinking plasma, leading to an extended advective column (panel c). At still higher accretion rates, a fully subsonic, shock-free magnetospheric flow may become possible (panel d). The fully subsonic shock-free solution, which might be viewed as a high-accretion-rate extension of the advective regime, is derived and tested numerically in \citetalias{shock-free}.

According to \citetalias{BS76}, the radiation-supported shock appears  slightly below a
 critical luminosity $L^{*}$,
 {expression for which was proposed by \citetalias{BS76} in the geometry of the `accretion curtain'}. Typically $L^* \lesssim 10^{37}$~erg s$^{-1}$.
 A corresponding dimensionless parameter was introduced :
$\epsBS = L^*/L = 2 l_0  c/(\varkappa_{\rm T} \dot M)$ (see their section 4.3),
where  $l_0$   is the curtain footprint length at the NS surface and $\varkappa_{\rm T}$ is the Thomson scattering cross-section per gram, $\varkappa_{\rm T}\simeq\sigma_{\rm T}/m_{\rm p}$.
 Formally,  if $\epsBS=1$, 
 the distance above the NS surface, where the incoming material is stopped by the radiation pressure, becomes equal to the thickness $\delta$ of the accretion curtain.

The case of $\epsBS \ge 1$ is called `the shock regime' by \citetalias{BS76} and refers to solutions where the energy flux $F_{\rm rad}$ vanishes at the bottom of the shock, and all the energy released in the column is radiated by its sides ({case b in }Fig.~\ref{fig:drawing}).  
\citetalias{BS76} also introduced  parameter $\betaBS=F_{\rm rad}\, A_\perp\, \rstar/(\dot MGM)$, which \citetalias{AL23} called the `advection parameter'.\footnote{{The index `BS' is dropped from $\beta$ in Fig.~\ref{fig:drawing}. The value $\betaBS$ is analytically calculated in the \citetalias{BS76} approach, but the same definition is used to calculate $\beta$ in simulations by \citetalias{AL23}.}} Here $A_\perp$ is the \glaccept{surface area of the bases of the two columns}, and $F_{\rm rad}$ takes into account the work by pressure forces.
For the critical accretion, when $\epsBS = 1$,  advection parameter $\betaBS=0$. 
{Above the critical accretion rate, the cooling efficiency gradually decreases as the luminosity increases, while the advective parameter becomes positive,  $0<\betaBS<1$ ({case c in} Fig.~\ref{fig:drawing}).}

A zero-advection shock 
is considered  in the models by \citet{Becker+1998, Becker+Wolff2007, Becker+2012,
West+2017b},
where a balance between the characteristic timescale for photon escape and the accretion timescale is required. The balance is represented by a dimensionless `loss parameter' 
$
\xib \equiv {\uppi r_0 m_p c}/ \left(\dot{M}\sqrt{ \sigma_{\parallel}   \sigma_{\perp}} \right)$. 
The column in these simulations is considered cylindrical with the radius of $r_0$, which implies $l_0 = 2\uppi r_0$ in the notation of \citetalias{BS76}.
The definition of $\xib$ involves the cross-sections along ($\sigma_\parallel$) and across ($\sigma_\perp \sim \sigma_{\rm T}$) the direction of the magnetic field~\citep[see, e.g., section 3.1 of][]{Becker+Wolff2007}, which are used to determine the corresponding optical paths in the conservation equations. The radius $r_0$ is that of a filled column at the NS surface.  
The heating-cooling balance is marked by a specific value of $\xib$ of order of 1. We point to the similarity between parameters $\xib$ and  $\epsBS$ and their dependence on the cross-section. 

Regardless of the specific geometry, when the accretion rate is so high that either dimensionless $\xib$ or $\epsBS$ are much less than $1$, adiabatic heating can not be compensated by radiative cooling. 
In such  high$-\dot M$ regimes the energy flux at the NS surface is non-zero. 
Therefore, we propose that the transition between regimes with
different bottom-boundary conditions for the energy flux occurs near
\begin{equation}
\label{eq.condition_advection}
    \epsilon \equiv \epsBS \,{\varkappa_{\rm T}/\varkappa}\sim 1\, .
\end{equation}
 Here $\varkappa$ is a representative effective opacity in the accretion
column, which may differ substantially from the Thomson opacity
$\varkappa_{\rm T} $ in a strong magnetic field.
Eq.~\eqref{eq.condition_advection} can be thought of as a limit on the accretion rate, below which the heating-cooling balance is possible in the column.

\begin{figure}
    \centering
    \includegraphics[width=1\linewidth]{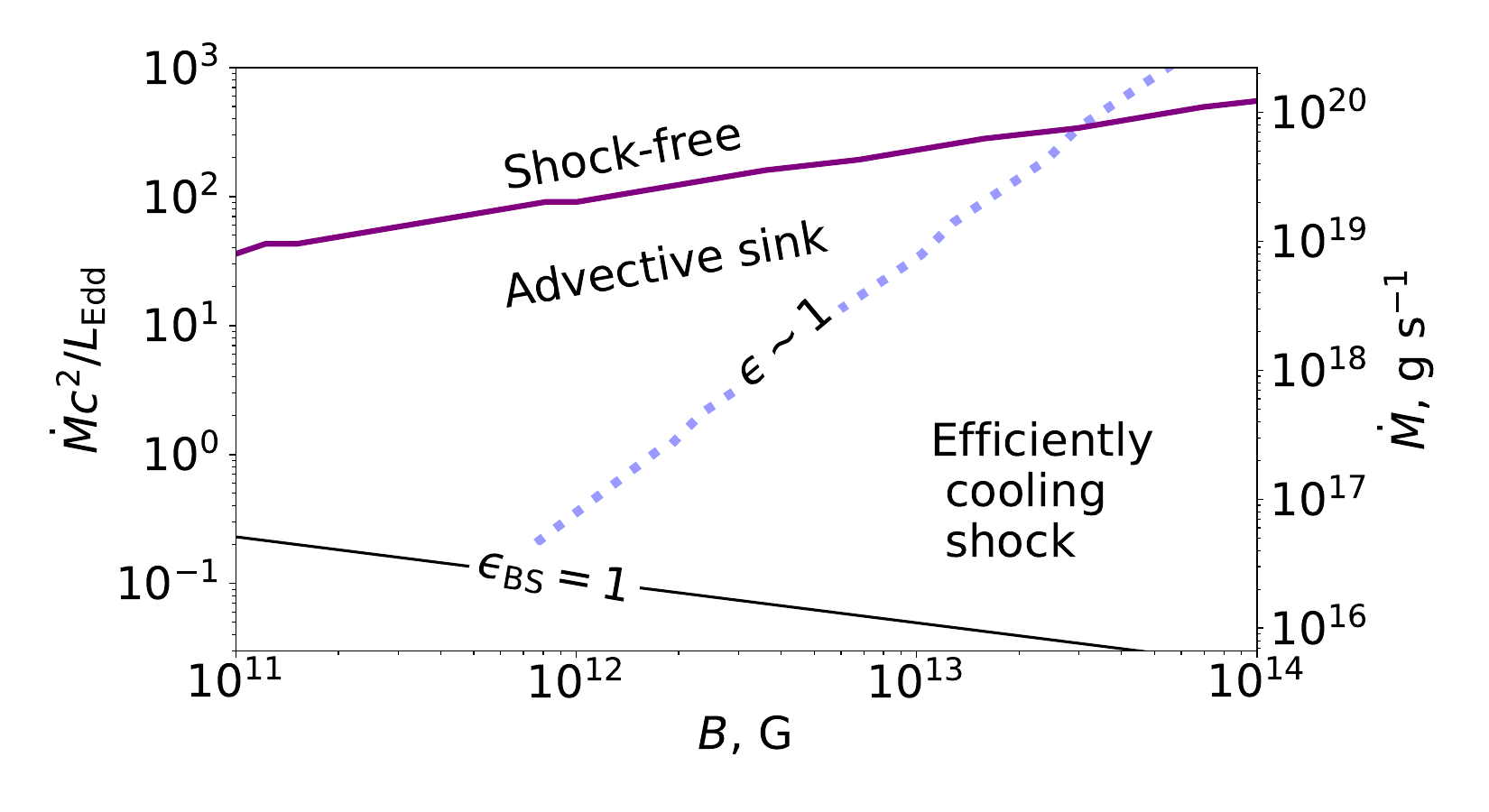}
   \caption{Magnetospheric accretion regimes.
   Radiation pressure becomes important  near the line $\epsBS=1$.   The shock-free solution limit is reproduced from results of \citetalias{shock-free}.  The dotted line is shown qualitatively, see Eq.~\eqref{eq.condition_advection}. 
   }
\label{fig:regimes_basic}
\end{figure}
   In Fig.~\ref{fig:regimes_basic}, we present the regimes of magnetospheric accretion{, which we considered above.}
   The lower solid line marks  the values corresponding to $\epsBS=1$.
To calculate it, we took the accretion curtain footprint length in the dipolar geometry $l_0 = 2\uppi \,a \,\rstar \sqrt{\rstar/\Rm}$~\citepalias{AL23}, where we set the NS radius $\rstar = 10^6$~cm
   and the magnetosphere radius $\Rm \equiv \ximag R_{\rm A} $ 
   with $\ximag=0.5$. The Alfv\'{e}n radius is defined as 
 \begin{equation}
 \label{eq.ra}
     R_{\rm A} = \left(  \frac{\mu^2} {2 \dot M \sqrt{2G\Mns}}\right)^{2/7}\, .
 \end{equation}
  The NS magnetic moment is $\mu = B\,\rstar^3/2$, where $B$ is the polar magnetic field.
The dimensionless factor $a$ is the azimuthal filling of the magnetosphere in one hemisphere and is fixed to $1/4$ {(the fiducial value used by \citetalias{AL23})}.
{The line `$\epsBS=1$' would be a transition between  (b) and (c) regimes (see Fig.~\ref{fig:drawing}) if the opacity  had not been modified in a magnetic field.}

The dotted line in Fig.~\ref{fig:regimes_basic} marked `$\epsilon\sim 1$' shows condition \eqref{eq.condition_advection},  which approximately takes into account opacity modifications.
We postpone the details of how this line is derived to Sect.~~\ref{sec:opacity_regimes}.  
Above this boundary the column cannot efficiently cool, and   there is a non-zero energy flux at the bottom of the column. In this case one should use advective solutions with $\beta > 0$. 
{Thus, condition \eqref{eq.condition_advection} approximately marks the transition between regimes (b) and (c), whose exact location, $\dot M(\mu)$,  depends on the microphysics of electron–photon interaction in a strong magnetic field.}

\section{Opacity in advective sinking columns}
\label{sec:opacity_bs}

Before discussing opacity-dependent transitions between accretion-column regimes, we first examine whether the Thomson-opacity approximation is self-consistent in the advective sinking solution of \citetalias{BS76}. The key quantity is the ratio between the characteristic photon energy inside the column and the local cyclotron energy~\citepalias[\cite{Canuto+1971};][]{BS76}. If the characteristic photon energy is below the cyclotron energy, the scattering opacity is expected to be reduced below the Thomson value.

We therefore investigate the range of magnetic moments and accretion rates for which the photon energy in the \citetalias{BS76} sinking solution is higher than the cyclotron energy. 
As the pressure at the bottom of the column is close to the critical magnetic pressure,
\begin{equation}
    \frac{a_{\rm rad} T_{\rm col, b}^4}{3} = \frac{B^2}{8\uppi},
\end{equation}
where $a_{\rm rad}$ is the radiation constant, the equilibrium temperature of the plasma and radiation is
\begin{equation}
    kT_{\rm col, b} \simeq 170\, B_{12}^{1/2}\ {\rm keV}.
    \label{eq:Tcrit_B}
\end{equation}
This temperature exceeds the cyclotron energy,
\begin{equation}\label{eq.Ecyc}
    E_{\rm cyc} \simeq 11.6\,B_{12}\ {\rm keV},
\end{equation}
for magnetic fields
\begin{equation}
\label{eq:max_B}
    B \lesssim 2.2 \times 10^{14}\ {\rm G}.
\end{equation}
This simple estimate already shows that, in sufficiently hot advective columns, the characteristic photon energies are  above the cyclotron energy.

\begin{figure}
    \centering
    \includegraphics[width=0.95\linewidth]{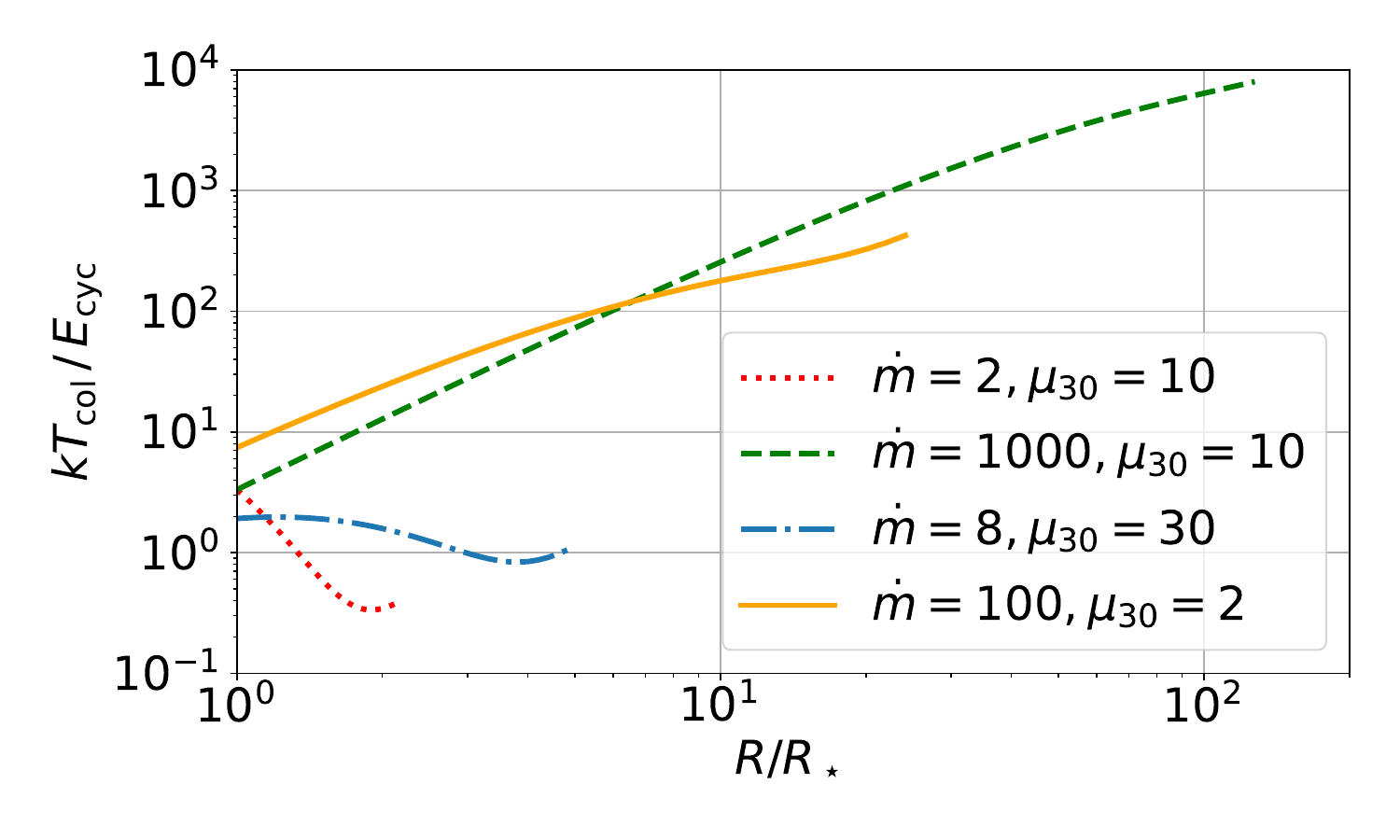}
    \caption{
    Profiles of the ratio $kT_{\rm col}/E_{\rm cyc}$ along the column height for different accretion rates and NS magnetic fields, calculated using the \citetalias{BS76} sinking solution. The adopted geometry has $\Delta \Re/\Re = 1/4$ and $a=1/4$.
    }
    \label{fig:profileTcol}
\end{figure}

For moderate accretion rates, the cyclotron energy may exceed the column temperature over part of the column. 
This is illustrated in Fig.~\ref{fig:profileTcol}, where we show profiles of $kT_{\rm col}/E_{\rm cyc}$ along the column height for several values of the accretion rate and magnetic field, calculated using the analytical \citetalias{BS76} sinking solution. If $kT_{\rm col}<E_{\rm cyc}$, the opacity can be reduced relative to the Thomson value, and the self-consistency of the Thomson-opacity \citetalias{BS76} solution becomes questionable. 
Hereafter, we generally assume that the magnetospheric flow occupies a fraction
$a=1/4$ of the azimuths.

\begin{figure}
    \centering
    \includegraphics[width=1\linewidth,trim={0cm 0cm 3cm 0cm},clip]{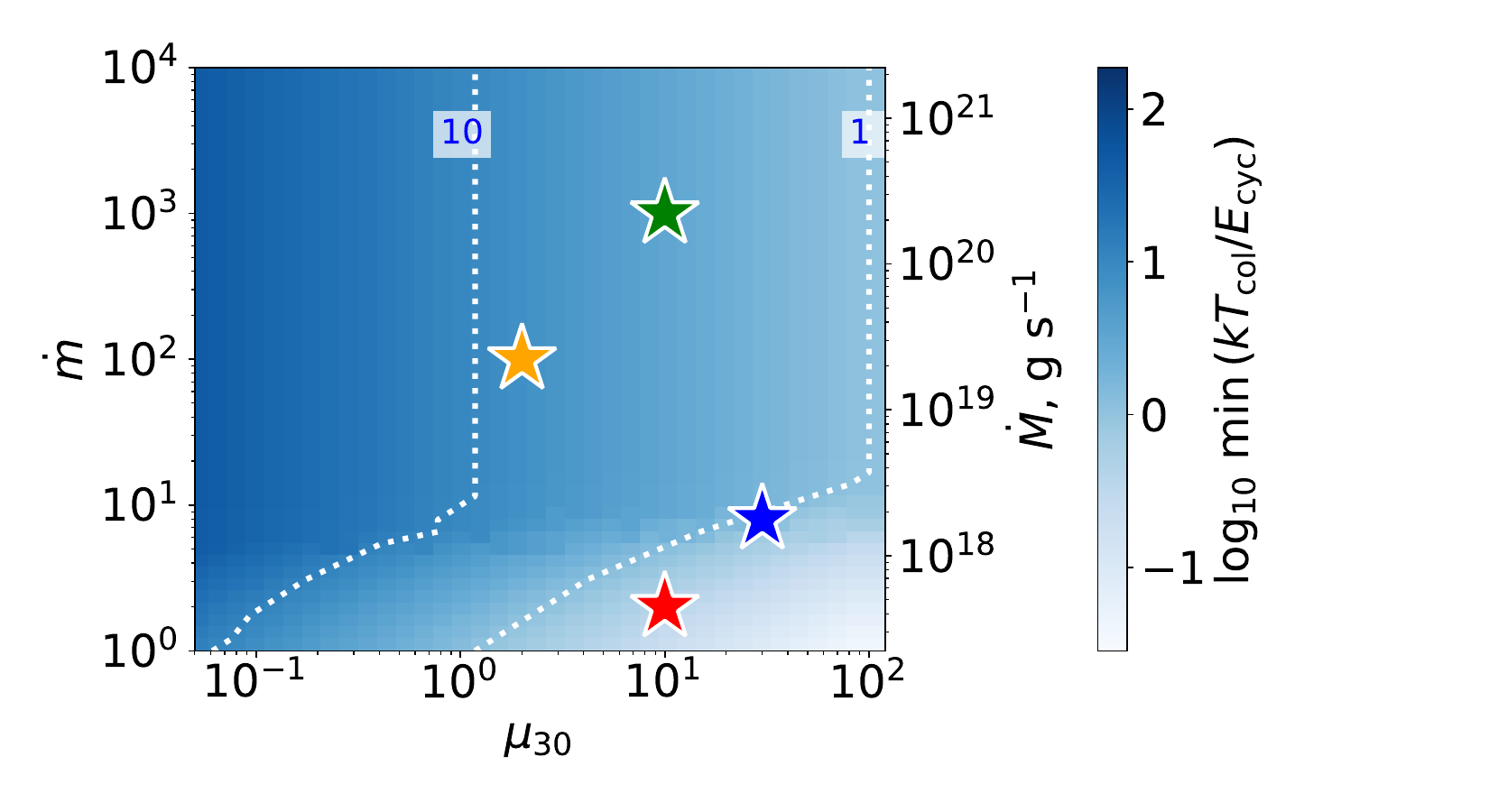}
    \includegraphics[width=1.1\linewidth]{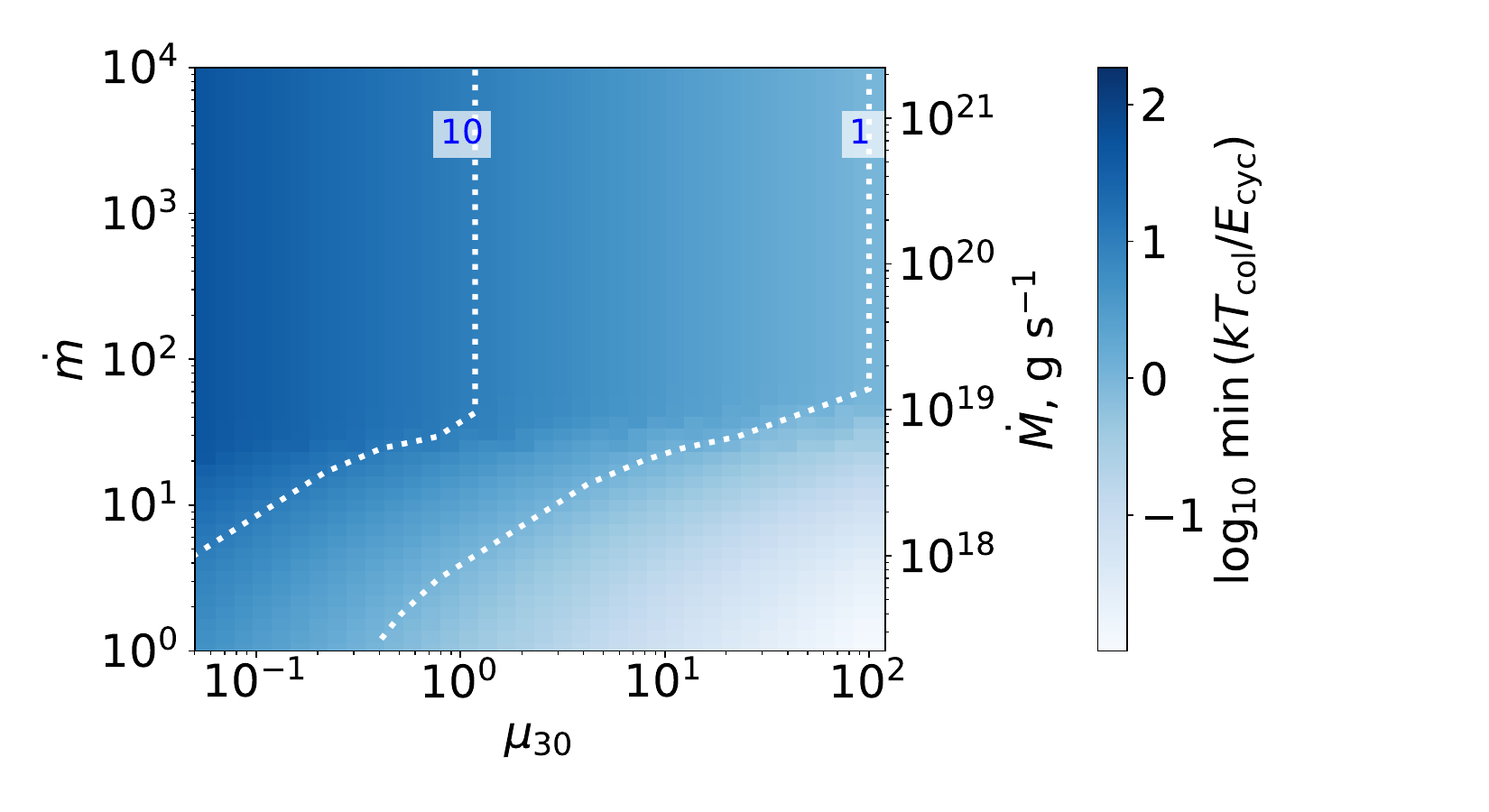}
    \caption{
    Minimum ratio of the equilibrium column temperature to the cyclotron energy, $\min(kT_{\rm col}/E_{\rm cyc})$, in the \citetalias{BS76} sinking solution. The upper panel corresponds to $\Delta \Re/\Re = 1/4$ and $a=1/4$, while the lower panel shows the case of full azimuthal filling, $a=1$. Dotted contours mark representative values of the ratio. Above the right contour, where $\min(kT_{\rm col}/E_{\rm cyc})=1$, the column temperature is everywhere high enough that the characteristic photon energy exceeds the cyclotron energy. Stars show the solutions in Fig.~\ref{fig:profileTcol}, their colour corresponding to the colours of the curves.
    }
    \label{fig:Tc_Ecyc}
\end{figure}

We next calculate the BS solutions over a grid of accretion rates and magnetic moments. 
For every solution with specific $\dot m$ and $\mu_{30}$, the minimum value of $kT_{\rm col}/E_{\rm cyc}$ within the column is found and the  result is shown in Fig.~\ref{fig:Tc_Ecyc} for two values of the azimuthal filling $a$. 
We utilize here normalized $\dot m = \dot M c^2/L_{\rm Edd} \approx \dot M / (2.2\times 10^{17}$\,g\,s$^{-1}$) and $\mu_{30}=\mu/(10^{30}$\,G\,cm$^3$). 
For every solution above the contour, where
$\min(kT_{\rm col}/E_{\rm cyc})=1$, the equilibrium temperature exceeds the local cyclotron energy everywhere within the column
Thus, in this part of the parameter space, the scattering opacity is not expected to be reduced below the Thomson value in the advective sinking solution. Depending on the location of the minimal ratio $kT_{\rm col}/E_{\rm cyc}$ within the column, the minimal values themselves depend differently on $\dot m$ and $\mu$. 
For larger mass accretion rates (growing curves in Fig.~\ref{fig:profileTcol}), the minimum is acquired near the surface of the star, and its value depends on the magnetic field only. In Fig.~\ref{fig:Tc_Ecyc}, this region is characterised by vertical contours of the minimal ratio.

For $a=1$, the range of accretion rates for which the sinking solution has $kT_{\rm col}<E_{\rm cyc}$ becomes larger (see the lower panel of Fig.~\ref{fig:Tc_Ecyc}). Hence, for a magnetospheric flow that fills all azimuths, efficiently cooling, moderate-height shock solutions are expected to remain possible over a wider range of accretion rates.

\begin{figure}
    \centering \includegraphics[width=1\linewidth]{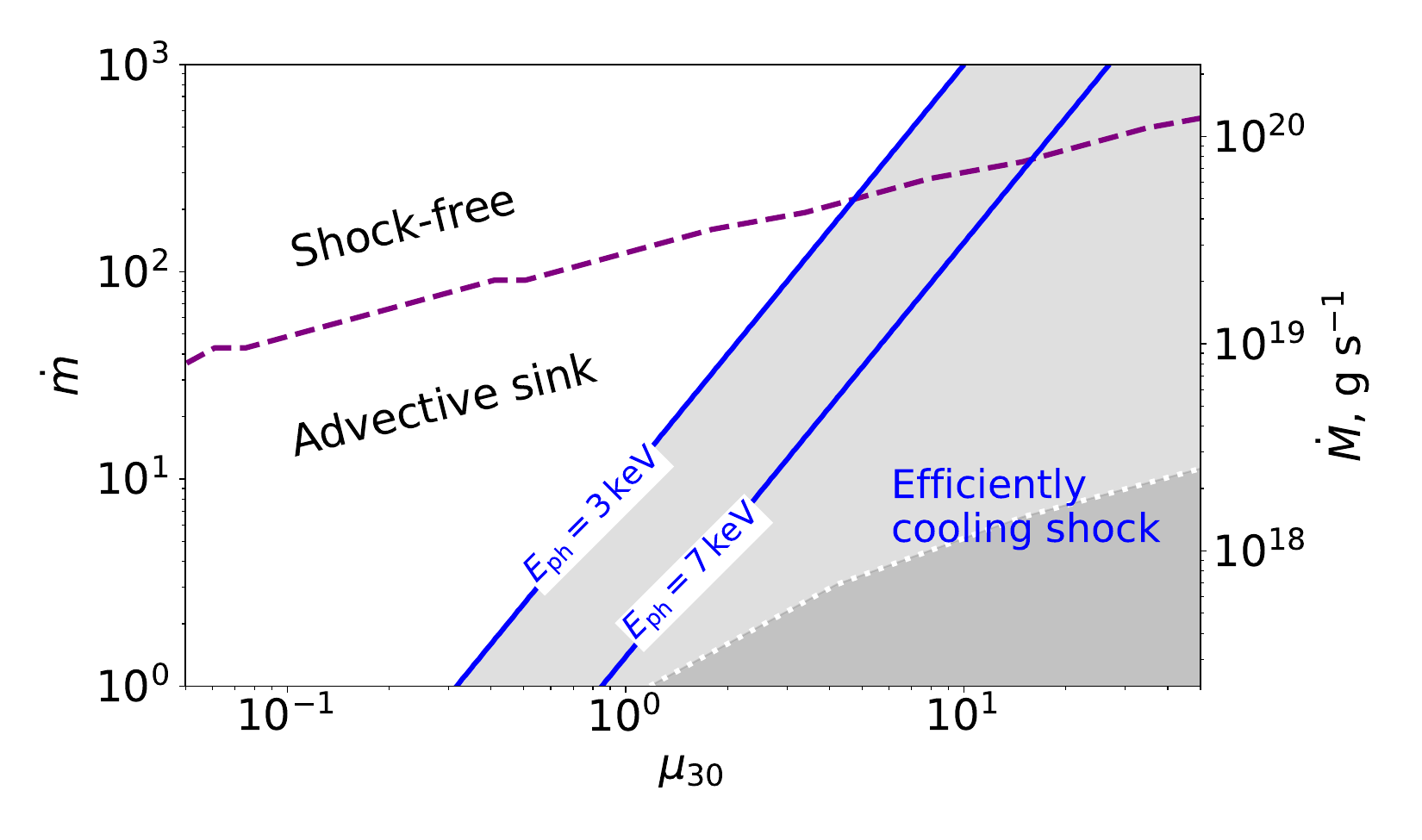}
 \includegraphics[width=1\linewidth]{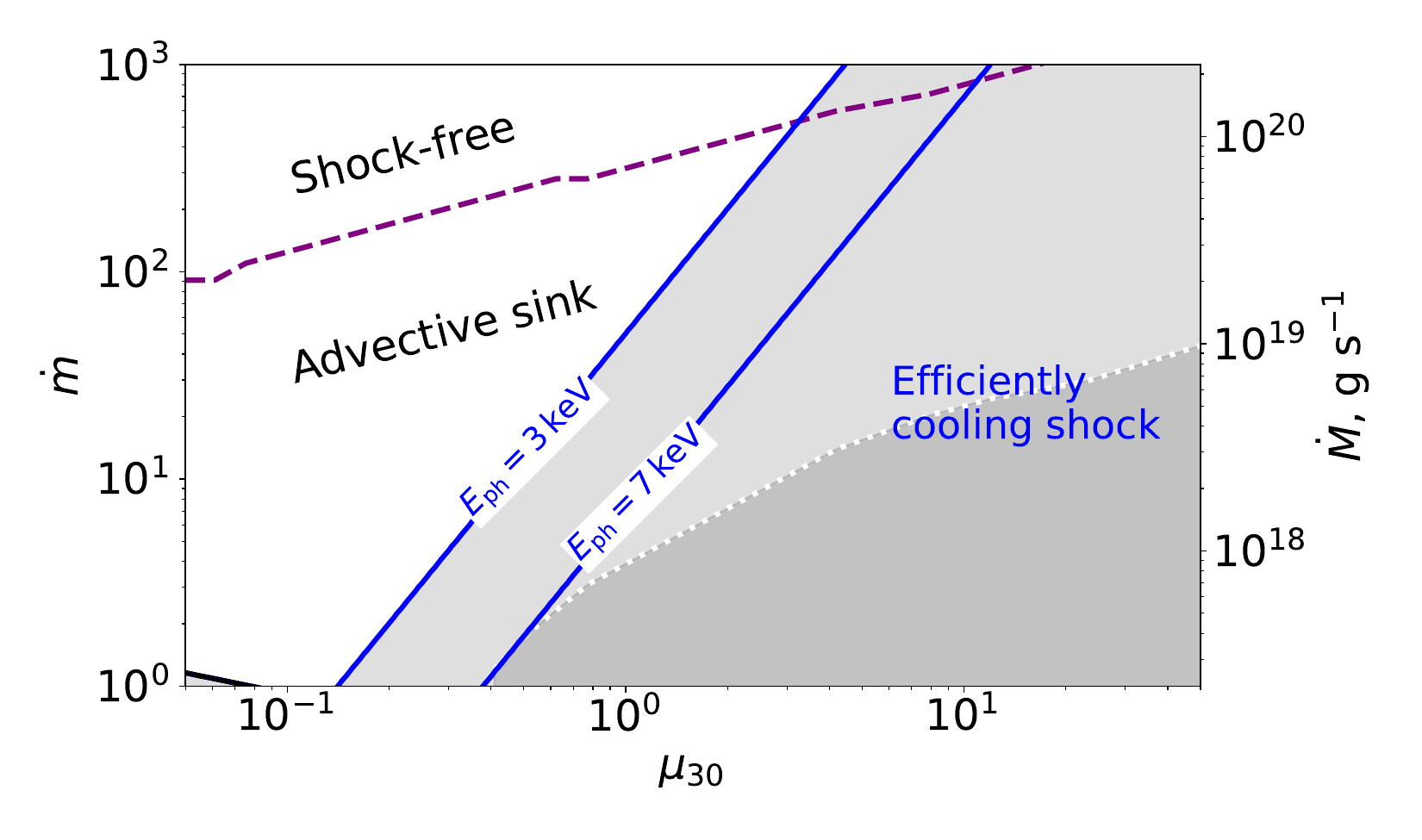}   
\caption{ Radiation-pressure dominated  magnetospheric accretion regimes. Above the blue lines only regime  with advection is possible for the  indicated representative photon energy. Below the blue lines, two regimes are possible: efficiently-cooling shock  and  advective sink. In the darkest grey area below the dotted line,  a replica from  Fig.~\ref{fig:Tc_Ecyc},   the characteristic energy of photons in the \citetalias{BS76} solution is below  $E_{\rm cyc}$.
Above the upper dashed line  the shock-free magnetospheric accretion is possible~\citepalias[see][]{shock-free}. 
Top and bottom: partial and full azimuthal filling of the magnetosphere, $a=0.25$ and $a=1$, respectively.  } 
  \label{fig:regims}
  \end{figure}

\section{Opacity-dependent regime boundaries}
\label{sec:opacity_regimes}

We now use the opacity considerations from Sect.~\ref{sec:opacity_bs} to estimate where different accretion-column regimes are expected. The central issue is whether an efficiently cooling column can maintain the balance between photon escape and gravitational energy release.
If this balance is impossible and the column keeps accumulating energy, the flow becomes advective.

In Sect.~\ref{sec.regimes}, we proposed a condition (Eq.~\ref{eq.condition_advection}) to separate  efficiently cooling columns from advective ones. 
For effective opacity $\epsilon \gtrsim 1$, radiative losses can balance the energy release in the column, allowing an efficiently cooling shock with $F_{\rm rad}=0$ at the stellar surface. For $\epsilon \lesssim 1$, cooling is insufficient and an advective solution with $\betaBS>0$ is required.

The value of $\epsilon$ depends on the opacity. As discussed in Sect.~\ref{sec:opacity_bs}, magnetic reduction of the scattering cross-section can allow efficiently cooling, relatively short accretion columns to persist to higher accretion rates. Once the characteristic photon energy becomes comparable to or larger than the local cyclotron energy, however, the opacity is no longer reduced.

As shown in Sect.~\ref{sec:opacity_bs}, advective sinking solutions are generally hot enough that the characteristic photon energy exceeds $E_{\rm cyc}$ over much of the relevant parameter space. Therefore, in the advective regime the scattering opacity is usually not expected to be strongly reduced below the Thomson value. The situation is different for efficiently cooling columns, where the characteristic photon energy can remain well below $E_{\rm cyc}$.

To estimate where the advective regime becomes unavoidable, we use Eq.~\eqref{eq.condition_advection} and, for illustrative purposes, we adopt the simplified opacity scaling for the photon energies below $E_{\rm cyc}$
\begin{equation}
\label{eq.kappy}
    \frac{\varkappa}{\varkappa_{\rm T}}
    \simeq
    \left(
    \frac{\overline{E}_{\rm ph}}{E_{\rm cyc}}
    \right)^2 ,
\end{equation}
where  $\overline{E}_{\rm ph}$ is the characteristic photon energy near the bottom of an efficiently cooling column
\citep{Wang-Frankl1981,Becker+Wolff2007}. In such columns, typical electron temperatures are expected to be in the range 1$-$10~keV for $B\sim 10^{12}-10^{13}$~G \citep{West+2017a}. Adopting two representative values of $\overline{E}_{\rm ph}$ and substituting \eqref{eq.Ecyc} and \eqref{eq.kappy} into \eqref{eq.condition_advection}, we plot in Fig.~\ref{fig:regims} the condition $\epsilon=1$ in blue.

Above each blue line in Fig.~\ref{fig:regims}, $\epsilon<1$ for the corresponding value of $\overline{E}_{\rm ph}$. In this region, an efficiently cooling shock solution is not possible, and the flow must become advective. 
{Conversely, below the blue lines, the temperature dependence of the opacity permits two possible stable solutions: a short, efficiently cooling column and an advective sinking solution, each corresponding to a different temperature.}
This creates a region of possible bistability, where two column configurations with substantially different heights may be realised for the same global parameters.
\begin{figure}
    \centering
    \includegraphics[width=\linewidth]
    {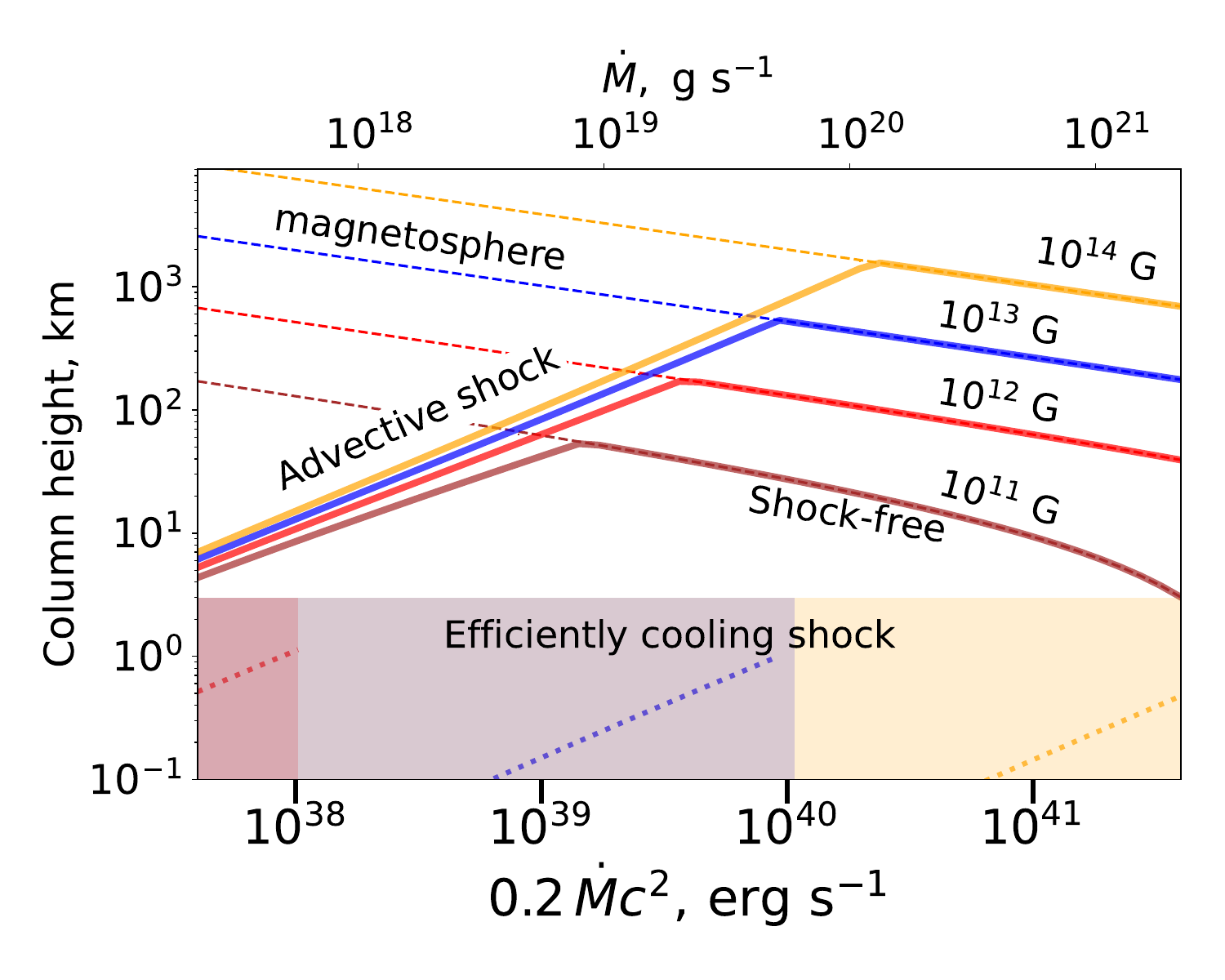}
    \caption{
    Bistability of  accretion-column solutions. Column height is shown in units of $\rstar$ as a function of accretion rate for different magnetic-field strengths.
    The solid rising lines follow the \citetalias{BS76} solution. The dashed lines show the magnetospheric radius, which limits the maximum column height. Shock-free solutions emerge to the right of the breaks. The shaded rectangles schematically indicate efficiently cooling shock solutions. Their maximum accretion rates correspond to the $3$~keV line in Fig.~\ref{fig:regims} (top). Dotted curves show the shock position estimated from the \citet{1988SvAL...14..390L} approximation, with the parameter $\gamma$ modified taking into account Eq.~\eqref{eq.kappy}. Each colour corresponds to a specific value of the magnetic field indicated along the solid curves. 
    }
    \label{fig:BSheights}
\end{figure}
The dotted line in Fig.~\ref{fig:regims} marks  the condition $\min(kT_{\rm col}/E_{\rm cyc})=1$ from Fig~\ref{fig:Tc_Ecyc}. Below it, in the dark shaded zone, the BS solution is not fully self-consistent even for 
$\varkappa = \varkappa_{\rm T}$.
In the case  of the full azimuthal filling, efficiently cooling columns can persist over a broader range of accretion rates. This is consistent with the temperature maps in Fig.~\ref{fig:Tc_Ecyc}, where the region with $kT_{\rm col}<E_{\rm cyc}$ is larger for $a=1$.

Corresponding column heights are depicted  in Fig.~\ref{fig:BSheights}.
The solid curves show the height of the accretion column predicted by the \citetalias{BS76} sinking solution for different magnetic-field strengths. The shaded horizontal rectangles schematically indicate efficiently cooling, moderate-height shock solutions. For example, for $B=10^{13}$~G, both a short efficiently cooling column and a tall advective column may be possible below the critical accretion rate indicated by the right edge of the blue shaded region.
For the efficiently cooling solutions, one can also plot the approximation given by \citet{1988SvAL...14..390L} {for the accretion shock height $R_{\rm s}$},
\begin{equation}
\displaystyle    \frac{R_{\rm s}}{\rstar}
    \simeq
    1+
    \frac{
    \ln\left[\eta\,\gamma^{1/4}(1+\gamma)\right]
    }{\gamma} \, . 
\end{equation}
Parameters $\gamma$ and $\eta$~\citep[see also][]{BS76} are modified to account for the magnetically reduced opacity. As $\eta \propto \varkappa^{1/4}$ and $\gamma\propto\varkappa^{-1}$, we use
\begin{equation}
\label{eq.gamma_eta_on_kappa}
    \gamma
    =
    \gamma_{\rm T}\,\frac{\varkappa_{\rm T}}{\varkappa} \qquad \mathrm{and} \qquad \eta
    =
    \eta_{\rm T}\,
    \left(\frac{\varkappa}{\varkappa_{\rm T}}\right)^{1/4}\,
\end{equation}
together with 
Eq.~\eqref{eq.kappy} for $\varkappa$.
Here $\gamma_{\rm T}$ and $\eta_{\rm T} $ are the values calculated for the Thomson opacity. The resulting curves are shown as dotted lines in Fig.~\ref{fig:BSheights}.
The dotted lines broadly agree with the trend found in numerical results of RRMHD simulations with temperature-dependent magnetic scattering by \citet{Sheng+2023}. They found that, for fixed accretion rate, increasing the magnetic field strength results in a shorter accretion column, while increasing the accretion
rate at high magnetic fields has the opposite effect. Such behaviour largely correspond to both dependences, shown by either solid or dotted lines in Fig.~\ref{fig:BSheights}.

\begin{figure*}
    \centering
\includegraphics[width=0.45\linewidth]{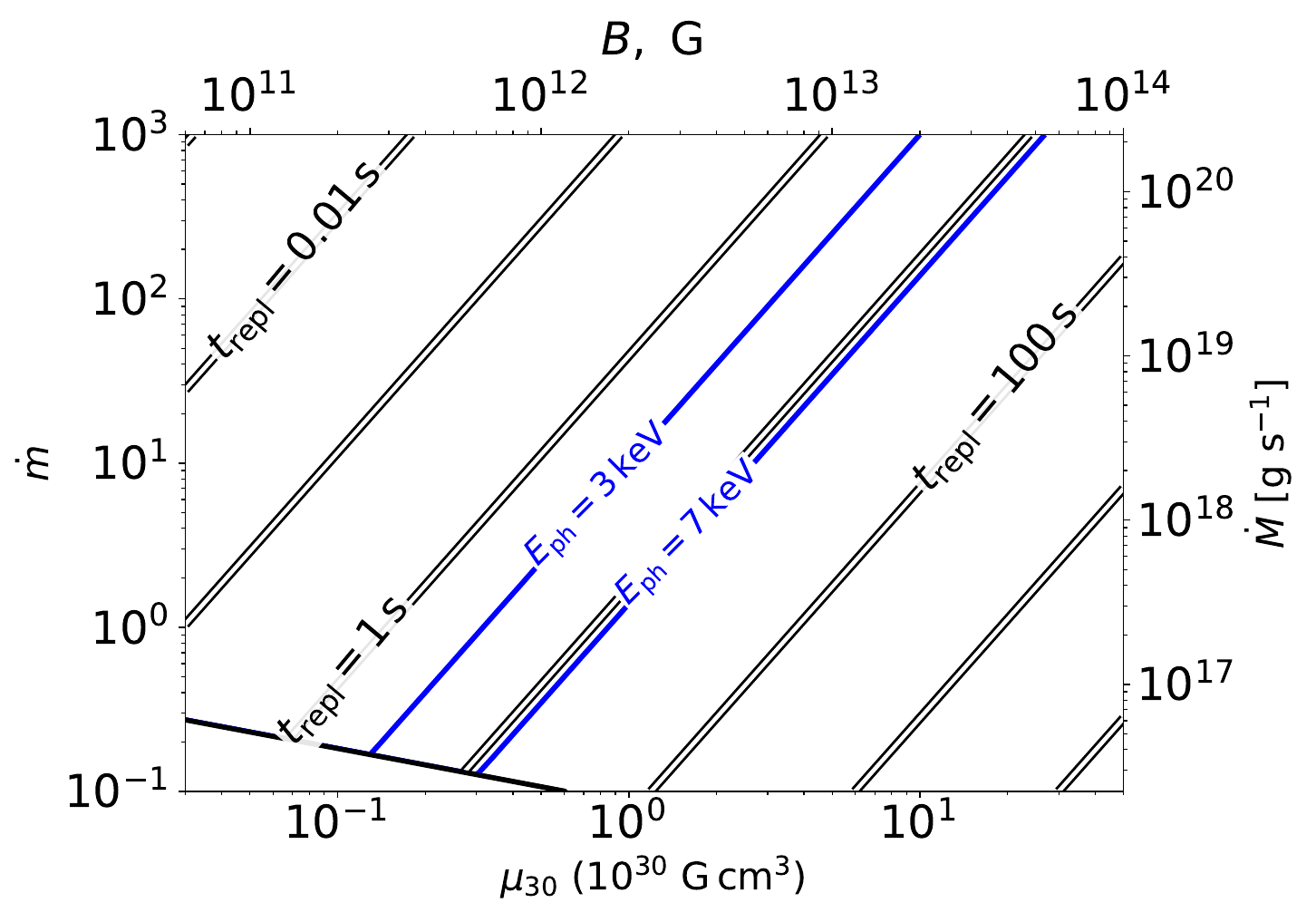}
\includegraphics[width=0.45\linewidth]{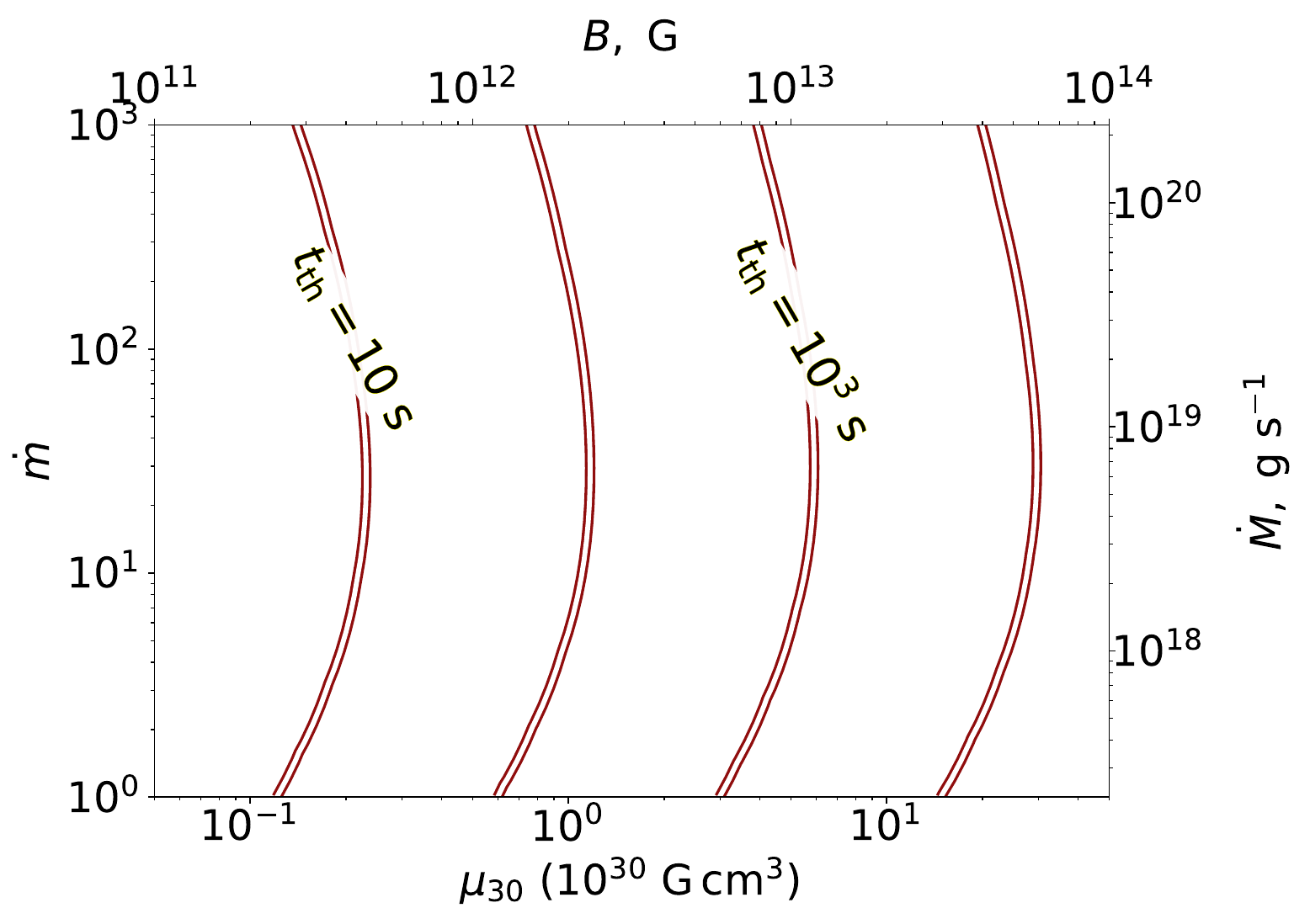}
    \includegraphics[width=0.45\linewidth]{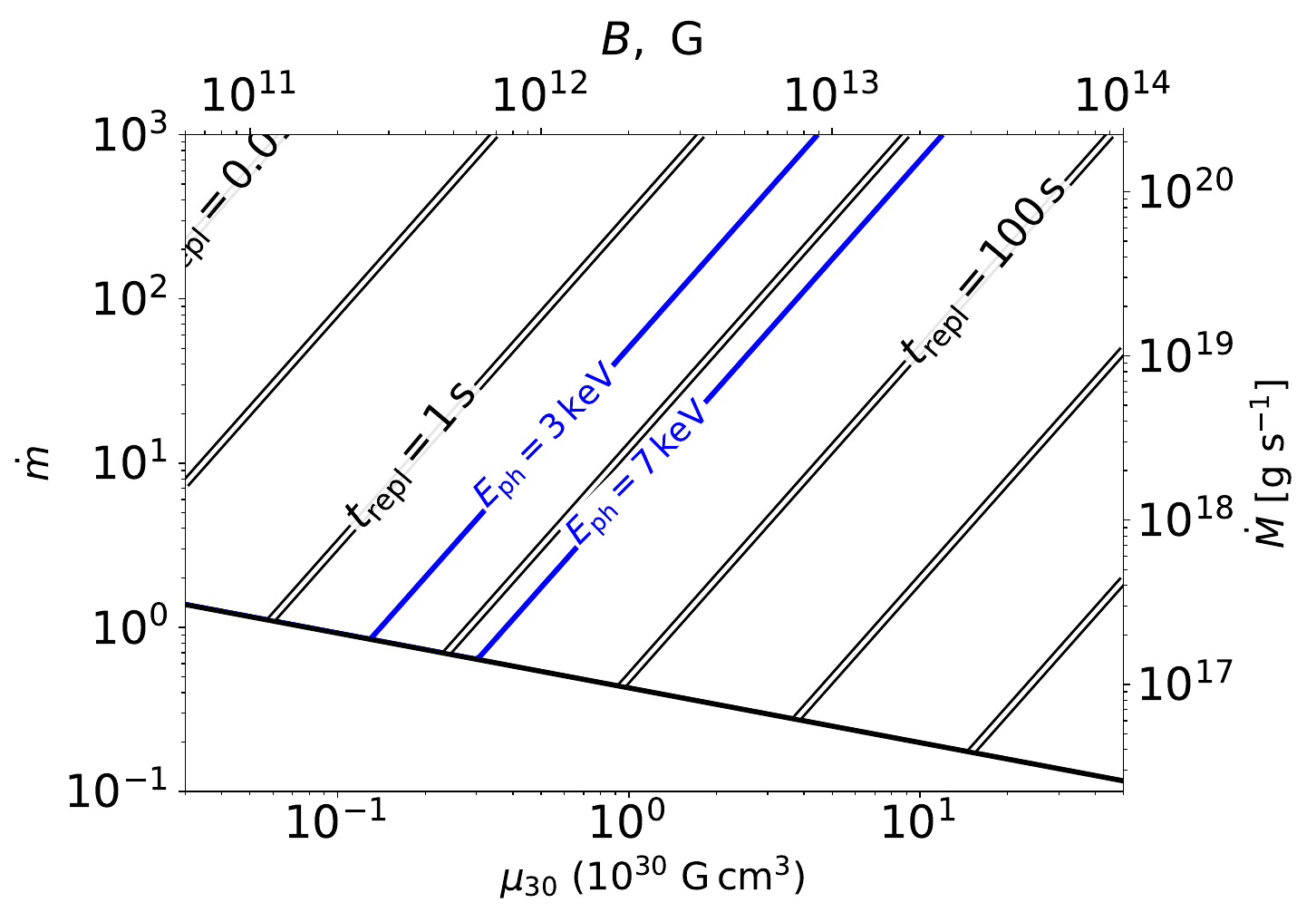}
    \includegraphics[width=0.45\linewidth]{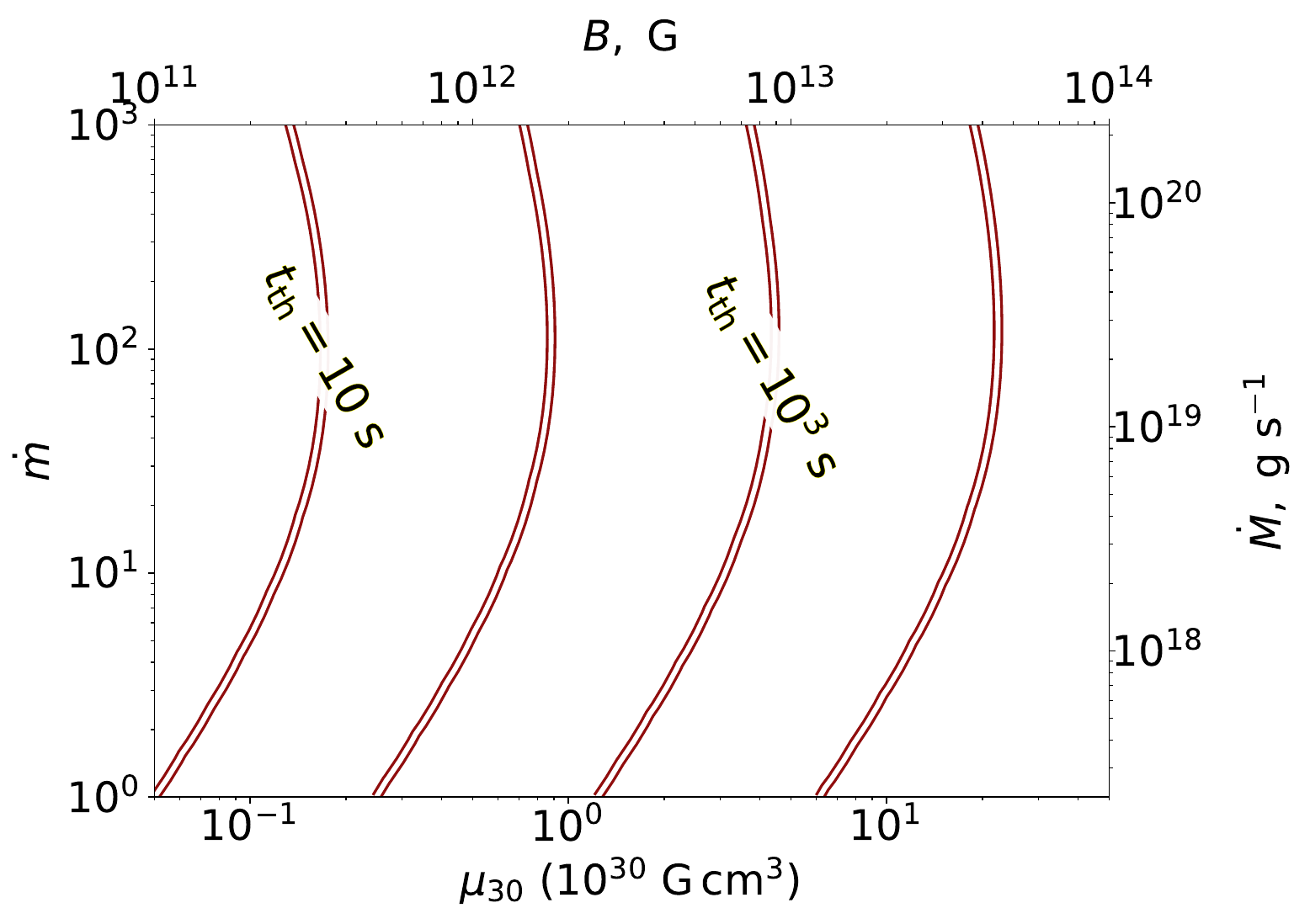}
    \caption{Replenishment (left) and cooling time (right) of advective column for $a=0.25$ (top) and $a=1$ (bottom).  The contours (double lines) separate one order of magnitude. The blue lines in the left panels repeat the critical accretion rates from Fig.~\ref{fig:regims}, and the black line corresponds to $\epsBS=1$.}
    \label{fig:times}
\end{figure*}
If the accretion rate increases during a transient event, the following sequence of events is possible. At low and moderate accretion rates, the source may remain in an efficiently cooling, short-column state. Once the critical value of $\dot M$ is reached, cooling becomes insufficient, and the column transitions to an advective sinking regime.
The column height can then increase substantially on a timescale comparable to the replenishment time $t_{\rm r}$, introduced by~\citetalias{AL23}. 
If the accretion rate increases further and the column height approaches the magnetospheric radius, the shock is likely to disappear, and the column switches to the shock-free regime introduced in \citetalias{shock-free}.
For sufficiently strong magnetic fields, such that the boundary between the advective sink and the shock-free solutions enters the shaded area in Fig.~\ref{fig:regims}, direct transition from a moderate-height shock to the shock-free regime is possible. 

In the parameter region of bistability, where solutions for efficiently cooling and advective columns coexist for the same global parameters, the column may undergo oscillations driven by the complex behaviour of magnetic opacities. { While this remains a qualitative prediction that has yet to be confirmed by dedicated time-dependent simulations, it points to a potentially rich  phenomenology in the accretion column dynamics.} Similarly, the blue lines in Figs.~\ref{fig:regims} and \ref{fig:BSheights}, which mark the transition between the two regimes, should be regarded as a first approximation to the true boundary, whose precise location will require more detailed opacity models. Nonetheless, they already capture the qualitative behaviour of this transition, and provide a useful benchmark for future, more accurate calculations based on realistic magnetic opacities~\citep[e.g.,][]{Suleimanov+2022}. In the Appendix, we compare  recent results of numerical simulations of the accretion columns with those predicted by theory.

\section{Discussion
} \label{sec:discussion}

The transition from an efficiently cooling column to an advective column is controlled by the competition between photon escape and advection. One useful timescale is the replenishment time~\citepalias[][]{AL23}
\begin{equation}
t_{\rm r} \simeq \frac{B^2\,A_{\perp} R_*^2}{8\pi\, G M\, \dot{M}} \,  ,
\end{equation}
where two symmetric accretion channels occupy surface $A_\perp$ on the NS surface. 
This timescale is relevant when an accretion column changes its vertical structure in response to a change in accretion rate or opacity. 
If the cooling balance is violated, the column 
{switches} to 
the advective state on a timescale comparable to the time required to replenish its mass. {Disbalance in mass and energy may become the driving force for the repeated transition between the states in a relaxation cycle with the period of the order $t_{\rm r}$. 
Such an concept was used, for instance, by \citet{2026arXiv260601625K} to explain the quasi-periodic oscillations of the X-ray pulsar 1A~0535+262.}
\del{Changes in column height are expected to affect the radiation pattern and spectrum. Such transitions may explain observed from X-ray pulsar 1A~0535+262 quasi-periodic hard X-ray flux variations 
as suggested by~\citet{2026arXiv260601625K}.}

Adopting the approximate value for the area, see \citetalias{AL23},
${A_{\perp}}/({4\pi R_*^2}) \approx ({a}/{2})\,({R_*}/R_{\rm e})\,({\Delta R_{\rm e}}/{R_{\rm e}})$, one arrives at 
\begin{equation} 
\label{eq:trepl_val}
t_{\rm r} \sim  7.6  \, B_{12}^{10/7}\dot M_{18}^{-5/7}\, a \,\frac{\Delta \Re}{0.25\Re}
\,{\rm s} \,  ,
\end{equation}
 where the inner disc radius $R_{\rm e}= 0.5 \,R_{\rm A}$, see Eq.~\eqref{eq.ra}. Magnetic field  and accretion rate are normalized here as $B_{12} = B / 10^{12}{\rm \, G}$ and $\dot M_{18} = \dot{M} / 10^{18} {\rm \, g\, s^{-1}}$, respectively.
Thickness of the magnetic channel $\Delta R_{\rm e}$ may depend on the disc structure and on $\dot M$, as suggested in the context of the replenishment time by~\citet{2026arXiv260601625K}. It is worth noting that in the simulations by \citetalias{AL23} the critical pressure at the bottom of the column is reached later than the  replenishment time, which is likely related to the growing vertical extent of the column. This can modify the direct link between estimate \eqref{eq:trepl_val} and oscillation frequency.

Fig.~\ref{fig:times} presents the map of replenishment times for $\Delta R_{\rm e}/R_{\rm e} =0.25$ and two values of $a$.  Within approximation \eqref{eq.kappy}, the critical accretion rate for $\epsilon=1$ for constant $E_{\rm ph}$  changes as $\propto B^2$ and is parallel to contours of constant $t_{\rm r}$. Assuming that $B=10^{12}$~G, or $\mu_{30} = 0.5$,  characteristic photon energy $\overline{E}_{\rm ph} = 3$~keV, one obtains an estimate for the critical accretion rate $\dot m \sim 3-10$  for $a=0.25-1$ or $\dot M \sim (5-30)\,\times 10^{17}$ g s$^{-1}$.  Corresponding replenishment time is about $3-5$~s, weakly depending on $a$.

The theoretical picture illustrated by Fig.~\ref{fig:regims} has also implications for pulsating ULXs. Existence, albeit transient, of a short accretion column and free-falling plasma inside the magnetosphere would indicate a strong NS magnetic field (if the pulsar parameters fall within the grey area).  
And vice versa,
for a normal pulsar field of  $B\sim10^{12}$~G, and a highly super-Eddington accretion rate, we do not expect a short column, basing on our Fig.~\ref{fig:BSheights}. 
If the cyclotron absorption feature is produced by the sides of the accretion column, as proposed by~\citet[][but see also \citealt{2013ApJ...777..115P}]{2023A&A...674L...2S}, a 
narrow cyclotron line 
should disappear at high mass accretion rates because its formation requires a relatively short accretion column as the emission site. 

Another temporal scale, which can be connected to (in)stability of accretion advective columns, is the thermal time (the heat diffusion time scale in the direction perpendicular to the field, see \citetalias[][equation 14]{AL23})
\begin{equation}
 t_{\text{th}} \simeq \frac{3\varkappa_{\rm T} \dot{M}}{c v} \frac{\delta^2}{A_\perp} \, ,
 \end{equation}
 where $\delta$ is the thickness of the `curtain' flow.
 For  \hbox{M51\,ULX$-$7}, \citet{2024A&A...689A.284I} reported detection of mHz-QPOs at super-Eddington luminosities with a property that when the QPOs were
detected, the spin pulsations could not be detected.  \citet{2010ApJ...710L.137F} found a QPO in the PDS of the PULX \hbox{M82\,X$-$2} at a frequency  $3-4$ mHz.
Observed characteristic frequencies are of order of the inverse thermal time in the sinking regime at the bottom of the column (see the right panels of Fig.~\ref{fig:times}), 
\begin{equation}
t_{\text{th}}
\sim 11 \, {B_{12}^{10/7} \, \left( \frac{\Delta \Re }{0.25\Re}\right)^2 \, \dot{M}_{18}^{2/7}}\,\beta_{\text{BS}}^{-1}~\text{s}\, .  
\end{equation}
(In the model of \citetalias{BS76}, $\betaBS$ is an implicit function of the other parameters, which lacks a good analytic approximation.)
Thus, the thermal-timescale variability for a super-Eddington NS with \(B \sim 10^{12}-10^{13}\,{\rm G}\) is expected to produce variability at frequencies of {$1-10$ mHz, consistent with observations of QPOs in \hbox{M51\,ULX$-$7} and \hbox{M82\,X$-$2}.} 
It is worth noting that the thermal time does not depend strongly on the accretion rate, which can explain why the frequencies of the detected mHz-QPOs hardly change between different epochs~\citep{2024A&A...689A.284I}.

{PULXs are known to show transient pulsations}. \citet{2020ApJ...891...44B} found no pulsations in several observations of M82\,X$-$2 of $\sim 60$~ks length.
\citet{2020ApJ...895...60R} reported vanishing of pulsations of \hbox{M51\,ULX$-$7} for $\sim 10^4$~s.
 Transient character of pulsations from ULX NGC 1313 X-2 was reported by \citet{2019MNRAS.488L..35S} in two out of six 30-ks time segments. 
Such time scales are longer than the replenishment or thermal times obtained by us and favour the disc-variations origin for the underlying evolution.

\section{Conclusions}\label{sec:conc}

Magnetospheric accretion can proceed through several regimes: hot spots, efficiently cooling radiative shocks, advective sinking columns, and fully subsonic shock-free flows.  The boundaries between these regimes depend on both macroscopic parameters and microscopic opacity.

   The transition between efficiently cooling and advective columns is controlled by the photon escape time. This can be expressed through a critical condition involving the effective opacity. If the effective opacity is reduced, efficient cooling can persist to higher accretion rates.
Strong magnetic fields favour short accretion columns and may allow strongly magnetized NSs to retain radiative shocks even at super-Eddington accretion rates.

In advective sinking columns, the plasma and radiation temperature near the base of the column rise comparing to efficiently cooling  columns.  For moderate magnetic fields and high accretion rates, the characteristic photon energy is typically above the cyclotron energy. 

Variable opacity can create a parameter region where both a short efficiently cooling shock and a tall advective column are possible. Transitions between these regimes may lead to changes in the column height, beam pattern, spectral shape, and pulse morphology.

     Precise regime boundaries require future calculations with realistic magnetic opacities, including polarization dependence, cyclotron resonances, Klein--Nishina corrections, and pair production.

\begin{acknowledgements} 
We are grateful to Ekaterina Sokolova-Lapa for the comments.
This work was funded by the Deutsche
Forschungsgemeinschaft (DFG, German Research Foundation), project
number \mbox{570950648} (GL), a grant from the Simons Foundation (00001470, PA), and the International Space Science Institute (ISSI) in Bern through ISSI International Team project \#495, ``Feeding the spinning top''.
\end{acknowledgements}

\bibliographystyle{aa}
\bibliography{mybib}

\appendix

\section{Predicted column heights in the sinking regime}

\begin{figure}
    \centering
    \includegraphics[width=\linewidth]
    {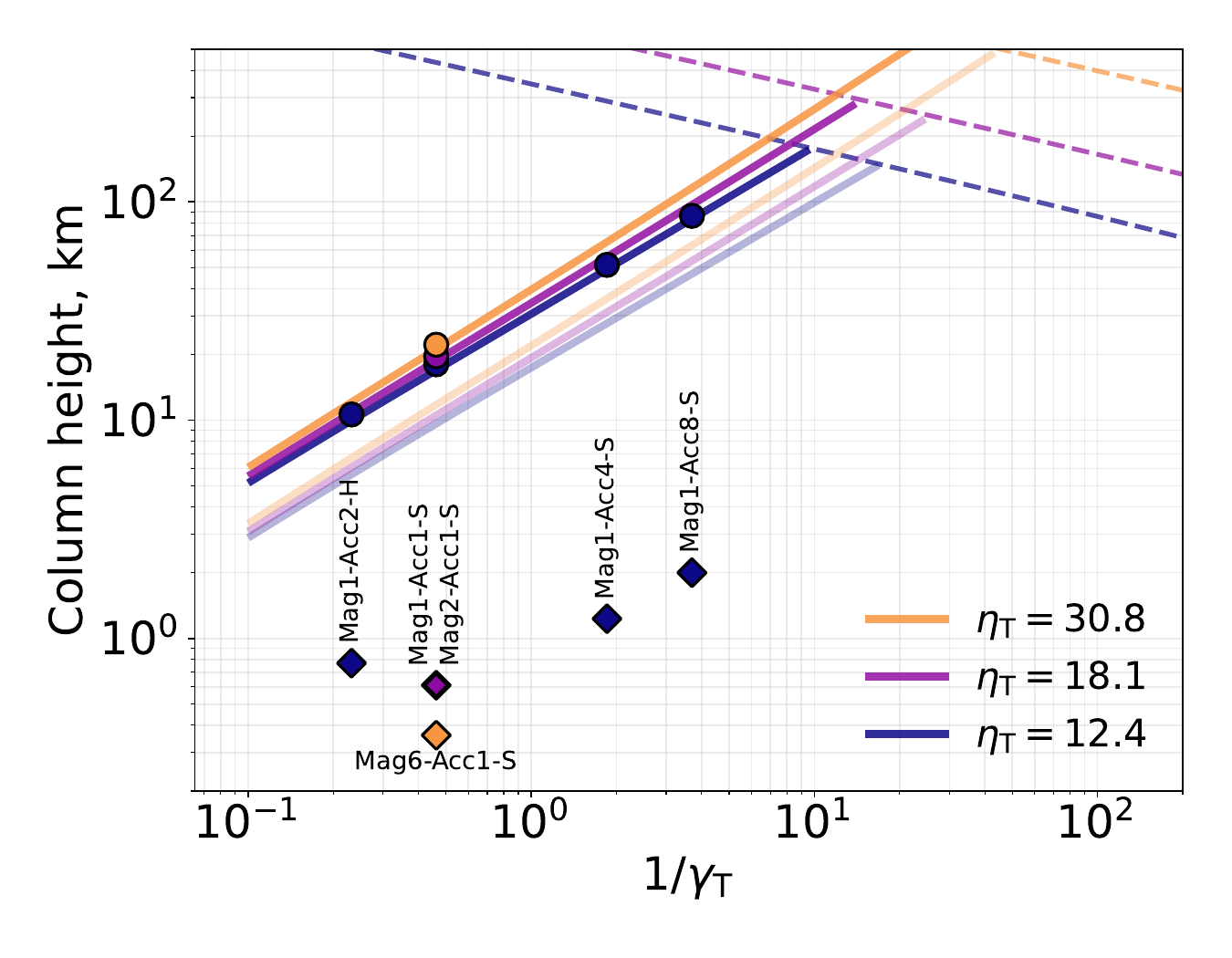}
    \caption{Modelled and predicted column heights for spherically diverging columns. The solid lines are the heights of advective columns in the sinking regime of \citetalias{BS76}. Diamonds with models' names show  values, obtained by \citet{Zhang+2025} in simulations, see their figure 11.    The circles mark the values, calculated from the parameters, as presented in their table 1. The transparent solid lines are the heights of advective columns in the sinking regime of \citetalias{BS76} for $\varkappa=0.5\,\varkappa_{\rm T}$. The dashed lines show the magnetospheric radii.
    }
    \label{fig:height_zhang_comp}
\end{figure}

RRMHD simulations by \citet{Zhang+2025}  for moderate magnetic fields ($(1-6)\times 10^{12}$~G) were performed with opacities from \citet{Suleimanov+2022}, which  accounted for strong magnetic fields and pair production. The resulting accretion columns heights are regularly smaller than predicted for the sinking regime in the model of \citep{BS76}, see Fig.~\ref{fig:height_zhang_comp}.

Using equations (34) and (36) from \citetalias{BS76}, we plotted in Fig.~\ref{fig:height_zhang_comp} the predicted height of the advective column as a function of $1/\gamma_{\rm T}$, which is proportional to the accretion rate, for different values of $\eta_{\rm T}$. This matches the presentation of figure 2 in  \citetalias{BS76}. The circles mark  predicted values of the column height,  calculated for the parameters from table 1 of \citet{Zhang+2025}.  We  took into account that their simulations assume a constant polar angle for magnetic lines, i.e., a spherically diverging flow. This case was also considered by \citetalias{BS76}. In the expressions for $\gamma_{\rm T}$ and $\eta_{\rm T}$  we  replaced  $d_0$ with the radius of the cone  $0.03\, R_{\rm NS}$.

In several simulations, the opacity at the column' bottom was reduced to about half of $\varkappa_{\rm T}$~\citep[see figure 4 of][]{Zhang+2025}. To show how the predicted heights vary with opacity, we plotted by the transparent lines the heights for $\varkappa=0.5\,\varkappa_{\rm T}$, taking into account dependences \eqref{eq.gamma_eta_on_kappa}. Apparently,  differences between the predicted and modelled values are still present.

\citet{Zhang+2025} argue that pair production  helps to cool the accretion column. At the same time, the opacity, which is expected to increase strongly due to pairs and  reduce cooling, is prescribed not to exceed $8\,\varkappa_{\rm T}   $ in their runs.  Moreover, their bottom boundary conditions allow a non-zero downward thermal flux.
Figure 6 of \citet{Zhang+2025} shows, that the radiation pressure  at the bottom of the column is about two order of magnitude smaller than $B^2/8\uppi$. Thus, the simulations, though achieved the force and thermal balance in most of the runs, did not approach the critical boundary conditions at the bottom of the column. This may explain the difference in column heights of the simulations and the sinking solution of \citetalias{BS76}. Is it a limitation of the particular simulations or a consequence of actual physics remains to be resolved.

\end{document}